\documentstyle[twocolumn,prl,aps,epsfig]{revtex} 
\begin{document}
\draft

\title{Entropy fluctuations for directed polymers in 2+1 dimensions}

\author{Xiao-Hong Wang$^{1,2}$, Shlomo Havlin$^{1}$ and Moshe
Schwartz$^{3}$}
\address{$^1$ Minerva Center and Department of Physics,
         Bar--Ilan University,
         Ramat Gan 52900, Israel}
\address{$^2$ Department of Thermal Science and Energy Engineering,
          University of Science and Technology of China,
          Hefei, Anhui 230026, P. R. China}
\address{$^3$ School of Physics and Astronomy, Raymond and Beversly
         Sackler Faculty of Exact Science, Tel-Aviv University, Tel-Aviv
         University, Tel-Aviv 69978, Israel}

\date{\today}

\maketitle
\begin{abstract}

    We find numerically that the sample to sample fluctuation of the
entropy, $ \Delta S $, is a tool more sensitive in distinguishing how from
high temperature behaviors, than the corresponding fluctuation in the free
energy. In 1+1 dimensions we find a single phase for all temperatures
since $ (\Delta S)^{2} $ is always extensive. In 2+1 dimensions we find a
behavior may look at first sight as a transition from a low temperature phase
where $(\Delta S)^{2} $ is extensive to a high temperature phase
where it is subextensive. This is observed in spite of the relatively large
system we use. The observed behavior is explained not as a phase transition
but as a strong crossover behavior. We use an analytical agreement to obtain
$ (\Delta S)^{2} $ for high temperature and find that while it is always
extensive it is also extremly small and the leading extensive part decays
very fast to zero as temperature is increased.

\end{abstract}


The problem of directed polymers has attracted much interest in recent 
years. It is relevant to many fields ranging from surface 
growth phenomena and spin glasses to flux lines in high-$ T_{c} $
superconductors [1-4]. It is well known that the problem of directed
polymers in a random medium is equivalent to the KPZ equation that
describes surface growth [5-7]. Much is known about the KPZ system, in
particular in 1+1 dimensions. The KPZ equation provides the exact
dynamical exponent for directed polymers in 1+1 dimensions [5-9]. The
situation for higher dimensions is more complex. 
Traditional approximation schemes like dynamical renormalization-group
methods fail to produce the exponents obtained by simulations [1,2,
10-15]. A self-consistent expansion of the correlation function
introduced later [16] yields results compatible with simulations for 2+1
dimensions. Above 2+1 dimensions, the behavior of directed polymers has a
phase transition when temperature is raised (in the directed polymer
problem or the level of noise in the KPZ system) and the system goes over
from a strong coupling to a perturbative weak coupling behavior [17-20].
All the field theoretical treatments agree that at 2+1 dimensions, namely
at the lower critical dimension itself, no transition should occur and the
strong coupling behavior exists at zero and any finite temperature
[16,19-21].  Some authors have claimed in the past on the basis of
numerical simulations, to have obtained a phase transition in 2+1
dimensional systems [22-28]. The system studied were, however, relatively
small and all those claims were not pursued eventually. In this
paper, we present numerical results obtained for much longer systems in
1+1 and 2+1 dimensions. We study the free energy fluctuations that is
usually studied in literature but in addition we obtain numerically also
the sample to sample fluctuations of
the entropy. This is a quantity, that is less common in the literature
and was introduced first, to the best of our knowledge, by Fisher and
Huse [29]. The fluctuations in the entropy are more pronounced than
those
of the free energy and in fact, we expect it to be extensive in the size
of the polymer in 1+1 and 2+1 dimensions. In higher dimensions we expect
a transition from a low temperature extensive phase to a high
temperature subextensive one. Namely defining the quantity
$ \lambda(t)=(\Delta S)^{2}/t $, its infinite volume limits, behaves like
an order parameter. It is zero at high temperatures and is of order 1 at
low temperatures. The numerical results concerning $ \lambda (t) $ 
in 1+1 dimensions are those being expected. We find that $ \lambda $ is of
order 1 over all the temperature region. In 2+1 dimensions, the numerical
results seem to indicate an unexpected transition from a low
temperature phase in which $ \lambda $ is of order 1 into a high
temperature regime where it is approaching zero. Although the transition
temperature 
seems to be defined very sharply and although $ \lambda (t) $ can be
fitted at high temperatures by $ \lambda (t) \propto lnt/t $, we claim
that it is just a crossover phenomenon and the reasons for it
occuring in such a spectacular way will be discussed later.
 
Consider a directed polymer on a hyperpyramid lattice structure with the
random energy assigned on each bond. The partition function $ G(R,t) $ for
directed polymers starting from $ (0,0) $ and ending at $ (R,t) $ is
defined by $ G(R,t)= \sum\limits_{C}e^{-E_{C}/T} $ where $ E_{C} $ is the
sum of the energy on the path $ C $ and $ T $ is the temperature. For
simplicity, we demonstrate our calculations using the transfer-martrix
method for the case of 1+1 dimensions. A similar formalism has been used
for 2+1 dimensions. The iteration relation for the partition
function $ G(R,t) $ is 
\begin{equation}
G(R,t+1)=G(R-1,t)e^{-\epsilon_{l}/T}+G(R+1,t)e^{-\epsilon_{r}/T},
\end{equation}
\noindent in which, $ {\epsilon}_{l} $ and $ {\epsilon}_{r} $ are the
energy assigned to the left and right bonds of the point $ (R,t). $ The
free energy $ F(t) $ is given by $ F(t)=-TlnG(t), $ where
$ G(t)=\sum\limits_{R}G(R,t) $ is the total partition function. The free
energy fluctuation $ \Delta F=(\overline{F^{2}}-\overline{F}^{2})^{1/2} $
has been commonly studied ($ \overline{A} $ is the ensemble average of
the quantity $ A $). We can also define the internal energy,
$ \langle E \rangle \equiv \sum\limits_{R}\sum\limits_{C}
E_{C}e^{-E_{C}/T}/\sum\limits_{R}\sum\limits_{C}e^{-E_{C}/T}. $ The 
internal energy fluctuation $ (\Delta E)_{T}=(\overline{{\langle E
\rangle}^{2}}-\overline{\langle E \rangle}^{2})^{1/2} $ is also an
interesting quantity. In order to obtain the iteration relation for the
internal energy $ \langle E \rangle $, we define $
\widehat{E}(R,t) \equiv
\sum\limits_{C(R,t)}E_{C(R,t)}e^{-E_{C(R,t)}/T}/G(t).$
It is clear that $ \langle E \rangle=\sum\limits_{R}\widehat{E}(R,t).
$ The iteration relation for $ \widehat{E}(R,t) $ is: 

\begin{center}
 $ \widehat{E}(R,t+1)=[e^{-{\epsilon}_{l}/T}\widehat{E}(R-1,t)G(t)  
 +e^{-{\epsilon}_{r}/T}\widehat{E}(R+1,t) $
\end{center}
\begin{equation}
G(t)+{\epsilon}_{l}e^{-{\epsilon}_{l}/T}G(R-1,t) 
+{\epsilon}_{r}e^{-{\epsilon}_{r}/T}G(R+1,t)]/G(t+1)
\end{equation}

To understand the difference between the free energy and
internal energy fluctuations, we use the entropy $ S = (\langle
E \rangle -F)/T. $ We present systematic numerical simulations for
the entropy fluctuation $ \Delta
S=(\overline{S^{2}}-\overline{S}^{2})^{1/2} $
based upon equations (1) and (2) with the initial conditions
$ G(R,0)={\delta}_{R,0} $ and $  \widehat{E}(R,0)=0. $ The random energy
assigned on the bond is assumed to be uniformly distributed in the
interval (-0.5,0.5) and uncorrelated in space and time. 
We use length up to $ t=2000 $ $ (d=1+1) $ and $ t=1000 $
$ (d=2+1) $ [30]. Six thousands configurations for 1+1 dimensions and
four thousands configurations for 2+1 dimensions were collected to take
the ensemble average. 

In 1+1 dimensions, the numerical results clearly show that the entropy  
fluctuation has the behavior $ (\Delta S)^{2} \propto t
$ for any temperature (see Fig. 1) [30]. It is expected that the entropy
fluctuation will tend to zero at the limits of zero and infinite
temperature. Indeed, the slope of $ (\Delta S)^{2}/t $ (Fig. 3(a)) is
about $ 1 $  for $ T<T_{P} $, i.e. $ \lambda (t) \sim T $ and about $
-4 $ for $ T>T_{P} $, i.e. $ \lambda (t) \sim T^{-4} $. As a result,
the free energy fluctuation and internal energy fluctuation will be the
same at the two limits. About $ T=T_{P}=0.2, $
the entropy fluctuation reaches a maximum. There is no evidence for a
phase transition in 1+1 dimensions.

The picture of directed polymers in 2+1 dimensions is more complicated
than that of 1+1 dimensions (see Fig. 2). Similar to 1+1 dimensions, the
entropy fluctuation tends to zero at the two limits of zero and infinite
temperature and there is a peak at $ T_{P}=0.11 $. For low temperatures
$ T \leq T_{P}=0.11, $  we see that the entropy fluctuation $ (\Delta
S)^{2} $ tends to $ t $ and $ (\Delta S)^{2} \sim T $ as in 1+1 dimensions
(see Fig. 2(a) and Fig. 3(b)). However, for $ T>T_{P}, $ we find that the
increase of entropy fluctuation as a function of $ t $  becomes slower and
slower as temperature is increased (see Fig. 2(b)). At the very high
temperatures, e.g. $ T=10.0 $, the entropy fluctuation is proportional to
$ lnt $ for large $ t $.
It seems that $ (\Delta S)^{2}/t $ will tend to the nonzero values only in
the region $ 0<T\leq T_{P}. $ The fluctuations of the free energy
$ \Delta F $ are
correlated with the fluctuations of the entropy $ \Delta S $. The high
temperature
behavior of $ \Delta F $ gives for large $ t $ a logarithmic dependence on
$ t $ while
for lower temperatures we see a crossover from logarithmic behaviour at
small $ t $ to $ \Delta F \sim t^{0.2} $ for larger $ t $'s. We have not
seen however, a sharp obvious temperature where this happens. In Fig. 3b
we see a sharp transition from a low temperature region $ T<T_{P} $,
where $ \lambda (t) $ is almost $ t $ independent to a high temperature
region where $ \lambda (t) $ is a decreasing function of $ t $. The
transition temperature is very close to the point where $ \lambda (T,t)
$ is maximal as a function of $ T $ for all $ t $.
  
    The observed behavior may be explained as follows. The infinite system
is characterized by a correlation length $ \xi $. As long as the size of the
system is adequately described in terms of the linear theory of deposition
proposed by Edwards and Wilkinson [31]. It is easy to show that within the
Edwards-Wilkinson theory $ (\Delta S)^{2} $ is proposional, for a finite system
to $ lnt $, as obtained by us for high temperatures. For $ t $ longer than
$ \xi $ the nonlinearities become important and $ (\Delta S)^{2} $ should be
extensive in $ t $. How is this related to the temperature dependence we
find? The correlation length $ \xi $ has a strong dependence on temperature,
$ \xi(T) $ is proposional to $ exp[(T/T_{0})^{\theta}] $
with $ \theta=3 $ according to Fisher and Huse [29] and $ \theta=2 $
according to Kim, Bray and Moore [18].
In any case the temperature dependence of $ \xi_{T} $ is so strong
that a relatively small increase in temperature may result in an increase
of $ \xi_{T} $ by orders of magnitude increasing it from values below
the minimal $ t $ we are using ($ t=50 $) to well above the maximal
value ($ t=1000 $). Indeed, a more careful examination of the data is
consistent with the above explanation.
In the temperature region $ T_{P}<T<2T_{P} $, all the lines but
the one correspondings to the smallest $ t $
merge. This suggest that the asymptotic value of $ \lambda (t) $
has already been reached, so that the asymptotic value is of order 1.
Thus above $ T_{P} $ the asymptotic value is still finite. This may explain
the former numerical results that claimed a phase transition in 2+1
dimensions [22-26]. The strong dependence of the correlation length on
temperature suggest that increasing the size of the system does not really
undergo a phase transition.

It is not difficult to obtain the asymptotic (infinite $ t $) high
temperature form of $ \lambda (T) $. As expected from the Edwards-Wilkinson
model $ \lambda (t) $ decrease as $ lnt/t $. This is true as long as
$ t>{\xi}_{T} $ but as $ t $ becomes of the order of $ {\xi}_{T}
\lambda $ must tend to a constant independent on $ t $. Therefore,
\begin{equation}
\lambda (T) \propto {\frac {ln{\xi}_{T}}{{\xi}_{T}}}
=({\frac {T}{T_{0}}})^{\theta}exp[-({\frac {T}{T_{0}}})^{\theta}].
\end{equation}
Thus we see that although $ \lambda $ is not zero it decreases extremely
fast with temperature.

The authors are grateful to Nehemia Schwartz and Ehud Perlsmann for useful
discussions. X.H.Wang acknowledges financial support from the Kort
Postdoctoral Program of Bar-Ilan University.



\vfill\break
\onecolumn

\begin{figure}
\epsfig{file=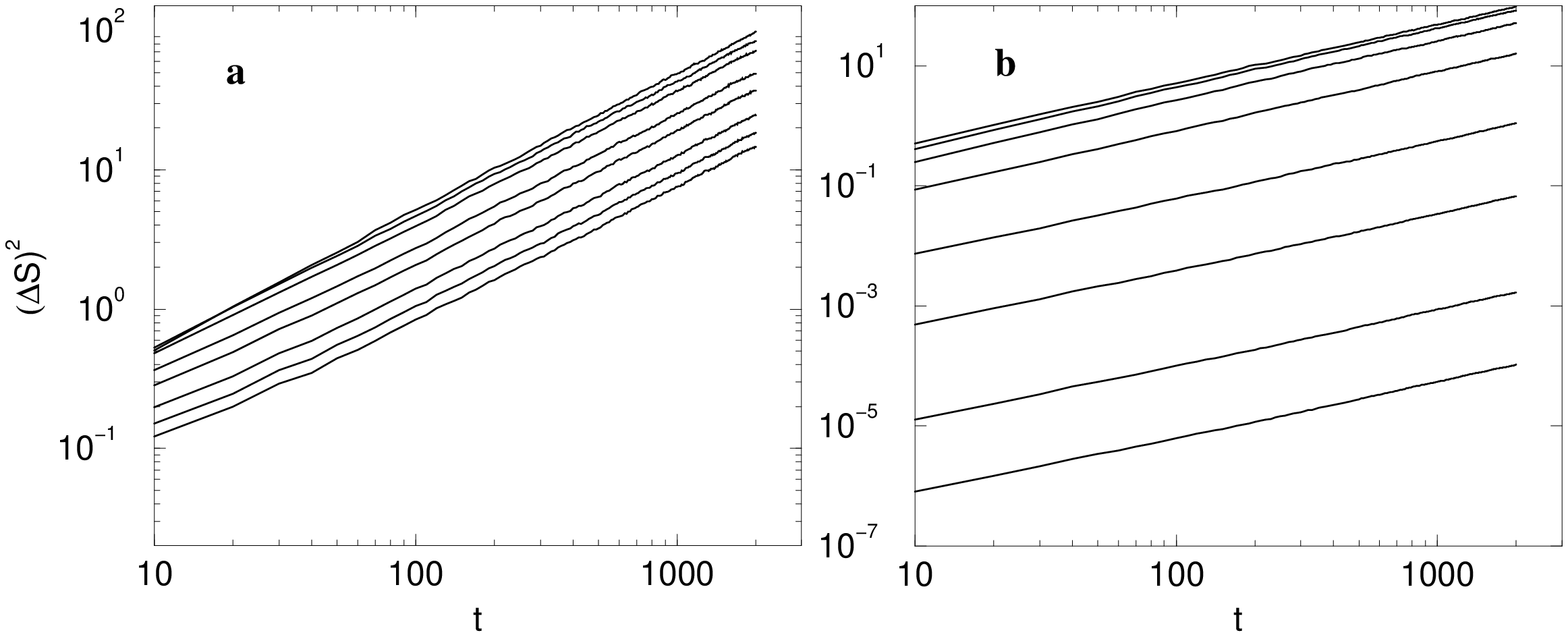,width=8cm,clip=,bbllx=0,bblly=0,bburx=350,
bbury=800,angle=-90}
\caption{ Plot of entropy fluctuation as a function of time $ t $ for
different temperatures in $ d=1+1 $, (a) for $ T={1\over 5}, {1\over 8},
{1\over 10}, {1\over 20}, {1\over 30}, {1\over 40}, {1\over 50} $ (from
top to bottom).
(b) for $ T=10, 5, 2, 1, {1\over 2}, {1\over 3}, {1\over 4}, {1\over
5} $ (from bottom to top).}
\end{figure}

\begin{figure}
\epsfig{file=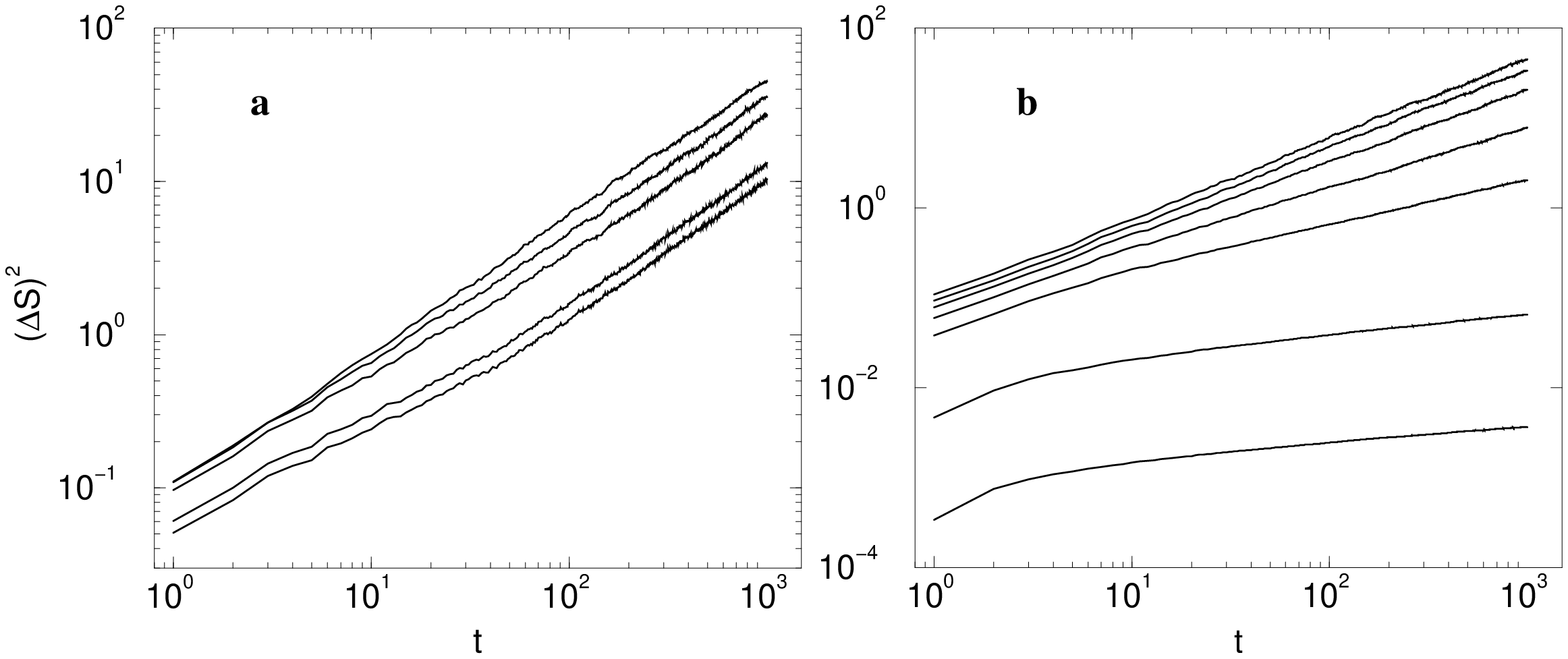,width=8cm,clip=,bbllx=0,bblly=0,bburx=350,
bbury=800,angle=-90}
\caption{ Plot of entropy fluctuation as a funtion of time $ t $ for
different temperatures in $ d=2+1 $, (a) for $ T={1\over 9.3},
{1\over 15}, {1\over 20}, {1\over 40}, {1\over 50} $ (from top to bottom).
(b)for $ T=1, {1\over 2}, {1\over 4}, {1\over 5}, {1\over 6}, {1\over 7},
{1\over 9.3} $ (from bottom to top) }
\end{figure}

\begin{figure}
\epsfig{file=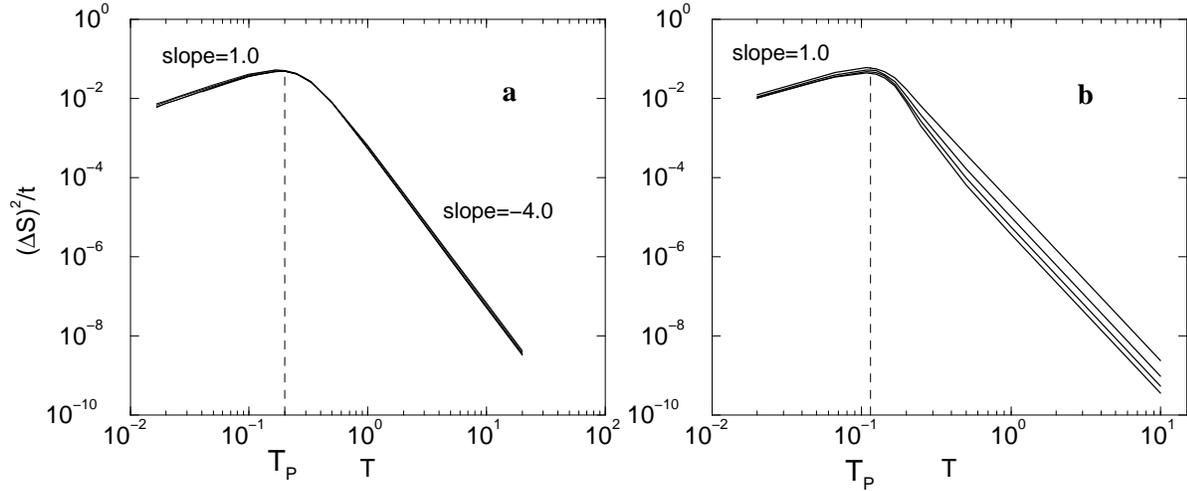,width=8cm,clip=,bbllx=0,bblly=0,bburx=350,
bbury=800,angle=-90}
\caption{ Plot of entropy fluctuation per unit length of the polymer as 
a function of temperature (a) for the different time $ t=50, 500, 1000,
2000 $ in 1+1 dimensions. (b) for the different time $ t=50, 100,
300, 600, 1000 $ (from top to bottom) in 2+1 dimensions } 
\end{figure}

\end{document}